\documentclass[sigconf]{acmart}

\AtBeginDocument{%
  }

\copyrightyear{2024}
\acmYear{2024}
\setcopyright{acmlicensed}\acmConference[CIKM '24]{Proceedings of the 33rd ACM International Conference on Information and Knowledge Management}{October 21--25, 2024}{Boise, ID, USA}
\acmBooktitle{Proceedings of the 33rd ACM International Conference on Information and Knowledge Management (CIKM '24), October 21--25, 2024, Boise, ID, USA}
\acmDOI{10.1145/3627673.3680012}
\acmISBN{979-8-4007-0436-9/24/10}

\usepackage{caption}
\usepackage{graphicx}
\usepackage{float} 
\usepackage{subfigure}
\usepackage{subcaption}

\begin{document}

\title{LSR-IGRU: Stock Trend Prediction Based on Long Short-Term Relationships and Improved GRU}

\author{Peng Zhu}
\authornote{These authors contributed equally to this work.}
\authornote{Corresponding author.}
\affiliation{
  \department{Department of Computer Science and Technology}
  \institution{Tongji University}
  \city{Shanghai}
  \country{China}}
\email{pengzhu@tongji.edu.cn}

\author{Yuante Li}
\authornotemark[1]
\affiliation{
  \department{Department of Computer Science and Technology}
  \institution{Tongji University}
\city{Shanghai}
  \country{China}}
\email{liyuante@tongji.edu.cn}

\author{Yifan Hu}
\affiliation{
  \department{Department of Computer Science and Technology}
  \institution{Tongji University}
  \city{Shanghai}
  \country{China}}
\email{2153592@tongji.edu.cn}

\author{Qinyuan Liu} 
\authornotemark[2]
\affiliation{
  \department{Department of Computer Science and Technology, Key Laboratory of Embedded System and Service Computing, Ministry of Education}
    \institution{
  Tongji University 
  \\ Artificial Intelligence Laboratory}
  \city{Shanghai}
  \country{China}}
\email{liuqy@tongji.edu.cn}

\author{Dawei Cheng}
\affiliation{
  \department{Department of Computer Science and Technology}
  \institution{Tongji University}
  \city{Shanghai}
  \country{China}}
\email{dcheng@tongji.edu.cn}

\author{Yuqi Liang}
\affiliation{
  \institution{Seek Data Group \\ Emoney Inc.}
  \city{Shanghai}
  \country{China}}
\email{roly.liang@seek-data.com}
\renewcommand{\shortauthors}{Peng Zhu et al.}

\begin{abstract}
Stock price prediction is a challenging problem in the field of finance and receives widespread attention. In recent years, with the rapid development of technologies such as deep learning and graph neural networks, more research methods have begun to focus on exploring the interrelationships between stocks. However, existing methods mostly focus on the short-term dynamic relationships of stocks and directly integrating relationship information with temporal information. They often overlook the complex nonlinear dynamic characteristics and potential higher-order interaction relationships among stocks in the stock market. Therefore, we propose a stock price trend prediction model named LSR-IGRU in this paper, which is based on long short-term stock relationships and an improved GRU input. Firstly, we construct a long short-term relationship matrix between stocks, where secondary industry information is employed for the first time to capture long-term relationships of stocks, and overnight price information is utilized to establish short-term relationships. Next, we improve the inputs of the GRU model at each step, enabling the model to more effectively integrate temporal information and long short-term relationship information, thereby significantly improving the accuracy of predicting stock trend changes. Finally, through extensive experiments on multiple datasets from stock markets in China and the United States, we validate the superiority of the proposed LSR-IGRU model over the current state-of-the-art baseline models. We also apply the proposed model to the algorithmic trading system of a financial company, achieving significantly higher cumulative portfolio returns compared to other baseline methods. Our sources are released at https://github.com/ZP1481616577/Baselines\_LSR-IGRU.
\end{abstract}



\maketitle

\section{Introduction}
The stock market is one of the important financial investment markets and has always been closely scrutinized by researchers, especially in stock price prediction. According to the Efficient Market Hypothesis's theoretical framework \cite{merello2019ensemble}, stock prices reflect all available information. However, the volatility of stock prices poses a significant challenge to accurate prediction of their trends \cite{west1988dividend}. In recent years, with the continuous development of cutting-edge technologies such as deep learning \cite{lecun2015deep} and graph neural networks~(GNN) \cite{wu2020comprehensive,cheng2020risk}, these methods have been widely applied to stock price prediction tasks to address this challenge. 

The traditional methods of stock prediction typically only consider the temporal information of stocks, using historical volume and price data to forecast future trends, which have yielded decent results \cite{dl1, NN-2}. However, price changes among stocks often exhibit correlations. For instance, stocks within the same industry or related sectors tend to be influenced similarly. This correlation presents new opportunities for enhancing the accuracy of stock prediction. Methods based on graph neural networks \cite{niu2020iconviz, zhu2022si} capture this correlation by exploring relationships between stocks, thereby improving prediction accuracy.
For example, \cite{THGNN} utilized historical price data to construct a daily corporate relationship graph, employing graph neural networks to learn and predict the correlations among different stocks, thereby assisting in stock price forecasting. \cite{MASTER} proposed a novel stock transformer model, which utilized historical data input into the network and employed a self-attention mechanism to capture the correlations among stocks. These methods, based on the GNN and transformer models, comprehensively considered the interrelation among stocks, thus providing greater advantages in stock price prediction tasks.

However, despite the potential of GNN in exploring stock relationships, their application often faces a series of challenges. Firstly, the vast amount of data in the stock market imposes high demands on the training and computational resources of GNN. Secondly, although existing methods have considered relationships between stocks, there are still challenges in refining and accurately modeling these relationships. For instance, even within the same industry, the degree of correlation between stocks from different sub-industries may vary, requiring more sophisticated models to capture this complex relationship.
In addition, current research primarily focuses on directly concatenating the temporal and relational representations of stocks and then using attention mechanisms for weighting and prediction. However, the relationships and temporal features among stocks are often intertwined and dynamically changing. These methods overlook the potential complex associations between the two representations of stocks. For example, the relationship on a particular day may affect future trading conditions.

To address these problems, this paper proposes a novel stock prediction model called LSR-IGRU.
Firstly, we utilize the secondary industry information of stocks to construct their long-term relationships.
Secondly, we compute the overnight price changes of any two stocks over a certain period and evaluate their correlation by calculating the cosine similarity between them, serving as the short-term relationships of the stocks.
Then, we incorporate the representations of long-term and short-term relationships into each step of the GRU model's input, allowing the long-term and short-term relationships of stocks to influence the representation and temporal features of the next day, thus better capturing the dynamic changes and correlations in the stock market.
Finally, we conduct comprehensive experiments on four different datasets.
The experimental results demonstrate that the method proposed in this paper outperforms the current state-of-the-art baseline models. Furthermore, we deploy the model to the real algorithmic trading platform of EMoney Inc., a leading financial services provider in China \footnote{https://www.emoney.cn/}. Across various portfolio optimization strategies, our model achieves significantly higher cumulative returns compared to existing models, confirming the effectiveness and feasibility of our method in practical applications.
In summary, the contributions of this paper are mainly reflected in the following aspects:
\begin{itemize}

\item We first utilize secondary industry information to construct a long-term relationship network between stocks, and combined overnight price information to build a short-term relationship network for stocks, offering new insights and methods for modeling stock relationship networks.

\item We improve the input at each step of the GRU model to enhance its modeling capability for temporal information and long short-term relationships, aiming to improve the accuracy of predicting stock trends.

\item We conduct extensive experiments on different datasets, demonstrating the superior performance and generalization ability of our method. Additionally, LSR-IGRU is characterized by its ease of training and deployment, which has been widely applied and evaluated on real-world trading platforms, further demonstrating its practicality and reliability.

\end{itemize}

\section{Related Work}
\subsection{Traditional and Machine Learning Methods}
Traditional methods such as Autoregressive~(AR) \cite{autoregressive}, ARIMA \cite{ARIMA}, and Exponential Smoothing \cite{exponsmoothing} have been widely used in stock prediction in the past, mainly for modeling linear trends.
With the advancement of computational technology, machine learning techniques such as Hidden Markov Models (HMM) \cite{HMM, HMM-base}, Support Vector Machines (SVM) \cite{SVM, SVM2}, Decision Trees \cite{DecisionTree}, and Neural Networks \cite{NN, NN-2, NN-3} have gained considerable attention. These methods are capable of effectively capturing complex nonlinear interactions. However, stock data are characterized by low signal-to-noise ratios, large trading volumes, frequent trading, significant price fluctuations, and influences from various factors. Consequently, one of the main challenges faced by these models is the tendency to overfit.

\subsection{Deep and Reinforcement Learning Methods}
With the rapid development of deep learning, 
recurrent neural network~(RNN) has demonstrated strong capabilities. They can effectively model long-term dependencies in time series data and utilize inputs such as stock price to forecast market trends \cite{akita2016deep, karim2022stock, LSTM-base}. In recent research, \cite{gupta2022stocknet} proposed the StockNet model based on GRU, which includes an injection module to prevent overfitting and a survey module for stock analysis. 
However, these deep learning models often exhibit instability when facing extreme market fluctuations \cite{farmer2012stock, he2020impact}. 
Therefore, models based on reinforcement learning have attracted attention due to their adaptability and continuous learning capabilities. 
For instance, \cite{AlphaStock} combined deep attention networks with reinforcement learning, optimizing parameters through discrete agent actions to maximize the Sharpe ratio of investments.
Reinforcement learning faces challenges such as the need for large data sets and difficulty in model interpretation, which can affect its use in financial markets.

\begin{figure*}[ht]
\centering
\includegraphics[width=0.95\textwidth]{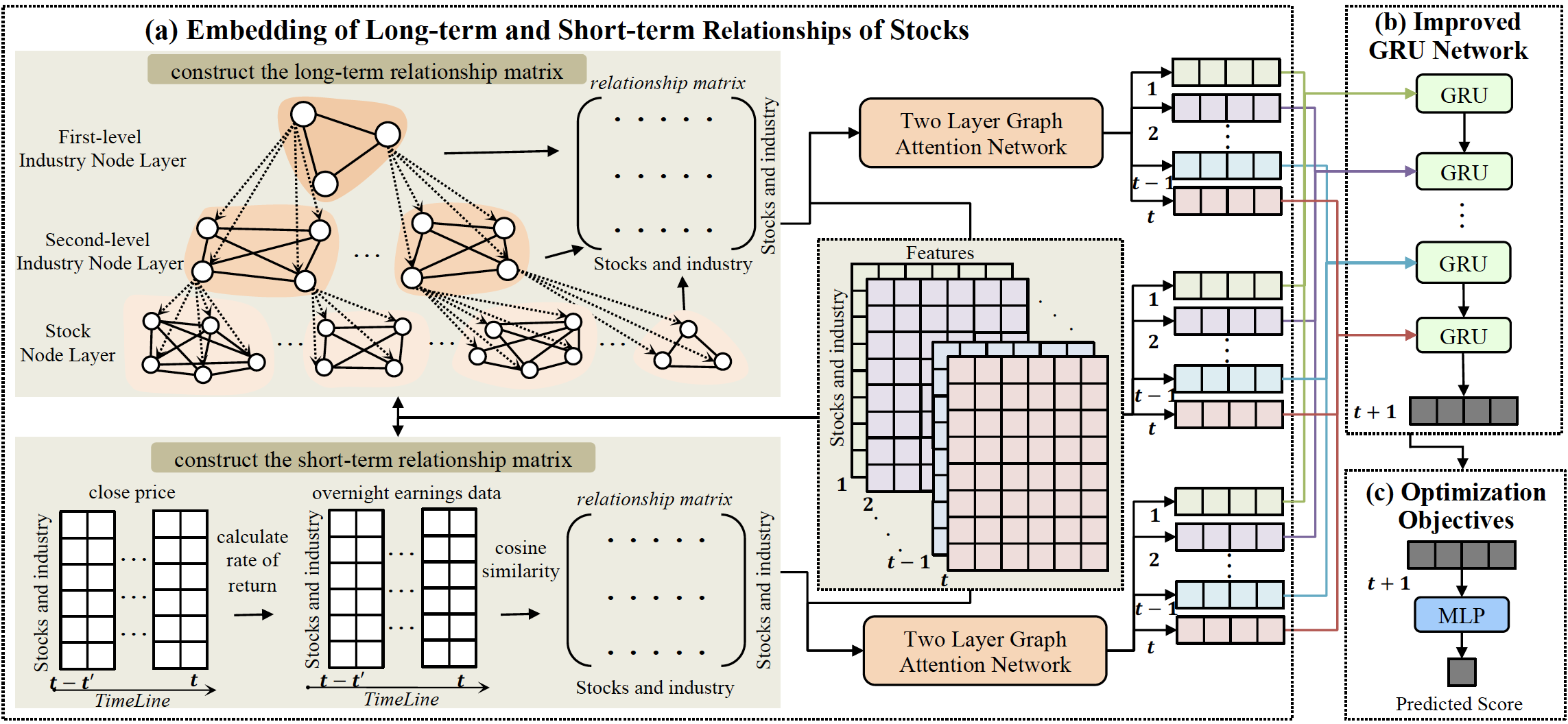} 
\caption{The architecture of the proposed model LST-IGRU. Part (a) outlines the generation of the long-short relationship representation model, which involves constructing long-term and short-term relationship matrices. These matrices are then inputted into respective GAT networks to derive relationship representations. Part (b) introduces enhancements to the GRU module, integrating both temporal and long-short relationship representations. Part (c) covers objective function optimization.}
\label{figure1}
\end{figure*}

\subsection{Graph Neural Networks and Latest Methods}
In recent years, GNN has been continuously applied to stock prediction because this technology can better capture the complex interdependencies in stock data. For instance, \cite{REST} proposed a hybrid model that combines RNN with GNN, achieving real-time predictions. \cite{THGNN} introduced a hierarchical attention mechanism into GNN to explore multi-level market dependencies, thereby enhancing the model's ability to conduct structured analysis of stock trends.
However, many graph-based models often overlook the diversity of stock price changes and the temporal nature of these changes, leading to the emergence of innovative graph-based models.
For example, \cite{MASTER} proposed a market-guided stock transformer capable of dynamically simulating the instantaneous and cross-temporal correlations of stocks, resulting in more accurate stock trend predictions. Nevertheless, despite showing certain advantages in stock prediction, GNN still has some shortcomings. These include inadequate modeling capabilities for complex nonlinear relationships and anomalous situations in the stock market, as well as insufficient robustness to data sparsity and noise.

Thus, our method captures both long-term trends and short-term fluctuations in stocks by integrating relational and temporal representations within each GRU step.

\section{The Proposed Model LSR-IGRU}
In this section, we provide a detailed exposition of the model LSR-IGRU, and the structure of the model is shown in Figure \ref{figure1}. 
\subsection{Predefined}
We define the set of all stocks as $S= \left \{s_{1},s_{2},\cdots,s_{m} \right \}$, where $s_{i}$ represents any stock and $m$ denotes the total number of stocks. For each stock $s_{i}$, its corresponding sets of primary and secondary industry codes are respectively defined as $Industry$ and $Industry'$, where the numbers of $Industry$ and $Industry'$ are $n$ and $n'$, respectively.
For any given stock $s_{i}$, its data on the $t$-th day is defined as $x_{it} = \left \{x_{it}^{open},x_{it}^{close},x_{it}^{high},x_{it}^{low},x_{it}^{volume},x_{it}^{turnover},x_{it}^{indtstry},x_{it}^{indtstry'} \right \}$,
where $x_{it}^{open}$, $x_{it}^{close}$, $x_{it}^{high}$,  $x_{it}^{low}$, $x_{it}^{volume}$ and $x_{it}^{turnover}$ denote the opening price, the closing price, the highest price, the lowest price, the trading volume and the trading amount respectively. $x_{it}^{indtstry} \in Industry$ and $x_{it}^{indtstry'} \in Industry'$ denote the primary industry code of the stock and the secondary industry code of the stock. 
The data of all days of stock $s_{i}$ is $x_{i}= \left \{x_{i1},x_{i2},\cdots, x_{it} \right \}$, and $t$ is the number of days of the stock. The data of all stocks is $X =  \left \{x_{1},x_{2},\cdots, x_{m} \right \}$, the data for all primary and secondary industries is the average of all stock data corresponding to that industry.
We define the long-term relationship matrix of stocks as $R_{long}$, and the short-term relationship matrix as $R_{short}$.

\subsection{Embedding of Long-term and Short-term Relationships of Stocks}
\subsubsection{Constructing the long-term relationship matrix}
First, we utilize the secondary industry information of stocks to construct the long-term relationships between them. We consider each stock's corresponding primary industry as a node in the graph because there exists a potential correlation among these industry nodes, thus connecting them. Each primary industry contains multiple sub-industries, which are regarded as secondary industries, and different secondary industries under the same primary industry correspond to different nodes in the graph, which are also interconnected. Under each secondary industry, there exist stocks corresponding to it, forming mutual connections between these stocks. By connecting stocks with stock nodes, stocks with industry nodes, and industries with industry nodes, we not only more accurately depict the long-term relationships among stocks in different industries, capturing potential correlations and common evolutionary trends in the stock market, but also reduce the parameter volume of relationship representation during graph model training.

From the previous definition part, we can know that the stock number is $m$, the number of primary and secondary industries are $n$ and $n'$ respectively, then the dimension $d = m+n+n'$ of the long-term relationship matrix $R_{long} \in \mathbb{R}^{d\times d}$ is expressed as follows:
\begin{equation}
 R_{long} =    
\begin{bmatrix}
  r_{11} &  r_{12} & \cdots &  r_{1d} \\
  r_{21} &  r_{22} & \cdots &  r_{2d} \\
  \vdots &  \vdots & \vdots & \vdots \\
  r_{d1} &  r_{d2} & \cdots &  r_{dd} \\
\end{bmatrix}
\end{equation}
where $r_{mm}=1$ indicates a connection between the two entities; otherwise, they are not connected. In addition, each row corresponds to the nodes of stocks, primary and secondary industries.

\subsubsection{Constructing the short-term relationship matrix}
Next, we conduct an in-depth study of overnight price fluctuations in stocks. A high opening in the morning suggests that investors generally believe the information from the previous day is favorable for the stock. Conversely, a low open indicates that the previous day's information may have unfavorable effects on the stock. If the overnight price fluctuations of two stocks are highly similar, it is likely that they are influenced by similar types of overnight information, demonstrating a higher level of correlation between them. Therefore, overnight price fluctuations in stocks can reflect their short-term relationships. Based on this theory, we calculate the overnight price fluctuations of any two stocks over a certain period and then evaluate the degree of correlation between them by computing their cosine similarity. Compared to correlation coefficients, cosine similarity better takes into account both the directional and magnitude synergies of overnight price changes, with its values ranging from -1 to 1. Through translation and scaling, the correlation between stocks can be transformed into an index ranging from 0 to 1.

We use any two stocks $s_{i}$ and $s_{j}$ to calculate the overnight changes over a period of time as the short-term relationship between two stocks and denote it as 
$Corr_{ij}=\left \{ {corr_{1},corr_{2},\cdots,corr_{t}} \right \}$, any $corr_{i}$ is expressed as :
\begin{equation}
\centering
\begin{split}
corr_{i}=\frac{\sum ([x_{i(t-t')}^{open},\cdots ,x_{it}^{open}])
([x_{j(t-t')}^{open},\cdots ,x_{jt}^{open}])}
{\sqrt{\sum {[x_{i(t-t')}^{open},\cdots ,x_{it}^{open}]^{2}}} \sqrt{\sum {[x_{j(t-t')}^{open},\cdots ,x_{jt}^{open}]^{2}}}}  \\
corr_{i} = (corr_{i}+1)/2
\end{split}
\centering
\end{equation}
where $t'$ represents the number of days looked back during calculation.
Using the same method, short-term relationships between stocks and industries, as well as between industries can be obtained. Then the dimension $d = m+n+n'$ of the long-term relationship matrix $R_{short} \in \mathbb{R}^{d\times d} $ is expressed as follows:
\begin{equation}
 R_{short} =    
 \begin{bmatrix}
  r_{11} &  r_{12} & \cdots &  r_{1d} \\
  r_{21} &  r_{22} & \cdots &  r_{2d} \\
  \vdots &  \vdots & \vdots & \vdots \\
  r_{d1} &  r_{d2} & \cdots &  r_{dd} \\
\end{bmatrix}
\end{equation}

\subsubsection{Using GAT to obtain representations of long-term and short-term relationships.}
A graph $G_{t}=(V, E)$ is constructed for any given day's stock data when obtaining long-term relationship representations of stocks and industries, where $V$ denotes the set of nodes and $V = \left \{s_{1},s_{2},\cdots,s_{m}, s_{m+1}, \cdots, s_{m+n}, s_{m+n+1}, \cdots, s_{m+n+n'} \right \}$, E denotes the set of edges and can be derived from the matrix $ R_{long}$. If $r_{ij} \in R_{long}$ is non-zero, it indicates the presence of an edge between nodes $s_{i}$ and $s_{j}$, m represents the number of stocks.
Firstly, for each node $s_{i}$, we compute the attention coefficient $\alpha_{ij}$ of its neighboring node $s_{j}$, indicating the importance of node $s_{i}$ to node $s_{j}$. This can be calculated as $e_{ij} = a(\rm W \left [x_{it}||x_{jt} \right ])$ and $ \alpha_{ij} = \frac{\exp(e_{ij})}{{\textstyle \sum_{k \in N_{i}}^{}} \exp(e_{ik})}$, where ${\rm W} \in \mathbb{R}^{2F} $ is the learnable weight matrix, $||$ represents the connection operation, $x_{it}  \in \mathbb{R}^{F}$ and $x_{jt} \in \mathbb{R}^{F}$ is the feature representation of nodes $s_{i}$ and $s_{j}$ in day $t$ respectively, $N_{i}$ represents the set of neighbor nodes of node $s_{j}$, $a$ is the mapping parameter, and $F$ is the dimension of the feature.

Then we aggregate the features of neighboring nodes of node $s_{i}$ using attention coefficients to generate a new representation $x_{it}^{long'}=\sigma ( {\textstyle \sum_{j\in N(i)}^{}}\alpha_{ij}x_{jt})$:
where $\sigma$ denotes a non-linear activation function ReLU. We use a 2-layer GAT in this model and follow the previous approach to obtain the representation $x_{it}^{long''} \in \mathbb{R}^{F'}$ after the two layers. For short-term relationship representations, we use the same two-layer GAT model for training and representation, denoted as $x_{it}^{short''}  \in \mathbb{R}^{F'}$, $F'$ is the dimension of the feature.

\begin{table*}[ht]
\centering
\caption{Comparing the experimental results of the models on four datasets. ARR measures the portfolio return rate of each predictive model, with higher values being better. AVol and MDD measure the investment risk of each predictive model, with lower absolute values being better. ASR, CR, and IR measure profits under unit risk, with higher values being better.}
\label{table1}
\footnotesize
\setlength{\tabcolsep}{0.2mm}{
\begin{tabular}{c|cccccc|cccccc|cccccc|cccccc}
\hline
Datasets & \multicolumn{6}{c|}{CSI 300}
         & \multicolumn{6}{c|}{CSI 500}
         & \multicolumn{6}{c|}{S\&P 500}
         & \multicolumn{6}{c}{NASDAQ 100}\\ 
Model & ARR $\uparrow$ & AVol $\downarrow$ & MDD $\downarrow$ & ASR $\uparrow$ & CR $\uparrow$ & IR $\uparrow$ 
      & ARR $\uparrow$ & AVol $\downarrow$ & MDD $\downarrow$ & ASR $\uparrow$ & CR $\uparrow$ & IR $\uparrow$ 
      & ARR $\uparrow$ & AVol $\downarrow$ & MDD $\downarrow$ & ASR $\uparrow$ & CR $\uparrow$ & IR $\uparrow$ 
      & ARR $\uparrow$ & AVol $\downarrow$ & MDD $\downarrow$ & ASR $\uparrow$ & CR $\uparrow$ & IR $\uparrow$ \\ \hline
BLSW        & -0.076 & 0.113 & -0.231 & -0.670 & -0.316 & 0.311  & 0.110  & 0.227 & -0.155 & 0.485  & 0.710  & 0.446 
            & 0.199  & 0.318 & -0.223 & 0.626  & 0.892  & 0.774  & 0.368  & 0.339 & -0.222 & 1.086  & 1.658  & 1.194 \\
CGM         & -0.185 & 0.204 & -0.293 & -0.907 & -0.631 & -0.935 & 0.015  & 0.229 & -0.179 & 0.066  & 0.084  & 0.001 
            & 0.099  & 0.250 & -0.139 & 0.396  & 0.712  & 0.584  & 0.116  & 0.242 & -0.145 & 0.479  & 0.800  & 0.603 \\
LSTM        & -0.214 & 0.175 & -0.275 & -1.361 & -0.779 & -1.492 & -0.008 & 0.159 & -0.172 & -0.047 & -0.044 & -0.128 
            & 0.142  & 0.162 & -0.178 & 0.877  & 0.798  & 0.929  & 0.247  & 0.176 & -0.128 & 1.403  & 1.930  & 1.386 \\
ALSTM       & -0.216 & 0.164 & -0.294 & -1.314 & -0.735 & -1.461 & 0.016  & 0.162 & -0.192 & 0.101  & 0.086  & 0.014 
            & 0.191  & 0.161 & -0.150 & 1.186  & 1.273  & 1.115  & 0.201  & 0.192 & -0.183 & 1.047  & 1.098  & 1.032 \\
GRU         & -0.229 & 0.156 & -0.290 & -1.469 & -0.790 & -1.631 & -0.004 & 0.159 & -0.193 & -0.028 & -0.023 & -0.118 
            & 0.324  & 0.169 & -0.139 & 1.917  & 2.331  & 1.657  & 0.225  & 0.188 & -0.165 & 1.197  & 1.364  & 1.160 \\
Transformer & -0.240 & 0.156 & -0.281 & -1.543 & -0.855 & -1.695 & 0.154  & 0.156 & -0.135 & 0.986  & 1.143  & 0.867 
            & 0.135  & 0.159 & -0.140 & 0.852  & 0.968  & 0.908  & 0.268  & 0.175 & -0.131 & 1.531  & 2.046  & 1.441 \\ \hline
TRA         & -0.074 & 0.169 & -0.222 & -0.436 & -0.332 & -0.409 & 0.125  & 0.162 & -0.145 & 0.776  & 0.866  & 0.657 
            & 0.184  & 0.166 & -0.158 & 1.114  & 1.172  & 1.106  & 0.267  & 0.181 & -0.144 & 1.475  & 1.854  & 1.427 \\
CTTS        & -0.193 & 0.206 & -0.312 & -0.937 & -0.618 & -0.907 & -0.041 & 0.172 & -0.239 & -0.241 & -0.173 & -0.237 
            & 0.154 & 0.161  & -0.113 & 0.952  & 1.356  & 0.965  & 0.349  & 0.197 & -0.193 & 1.769  & 1.808  & 1.610 \\ 
A2C         & -0.207 & 0.092 & -0.259 & -2.255 & -0.803 & -2.490 & -0.172 & 0.084 & -0.208 & -2.043 & -0.826 & -2.207 
            & 0.160  & 0.126 & -0.084 & 1.267  & 1.907  & 1.244  & 0.109  & 0.134 & -0.114 & 0.816  & 0.957  & 0.844 \\
DDPG        & -0.137 & 0.138 & -0.240 & -0.992 & -0.568 & -1.002 & -0.128 & 0.082 & -0.170 & -1.563 & -0.756 & -1.639
            & 0.111  & 0.129 & -0.091 & 0.864  & 1.223  & 0.887  & 0.130  & 0.156 & -0.131 & 0.832  & 0.994  & 0.863 \\
PPO         & -0.096 & \textbf{0.045} & \textbf{-0.120} & -2.138 & -0.800 & -2.234 
            & -0.032 & \textbf{0.015} & \textbf{-0.040} & -2.041 & -0.787 & -2.075 
            & 0.020  & \textbf{0.089} & \textbf{-0.067} & 0.220  & 0.291  & 0.263 
            & 0.148  & \textbf{0.118} & \textbf{-0.104} & 1.259  & 1.424  & 1.237 \\
TD3         & -0.154 & 0.137 & -0.252 & -1.122 & -0.610 & -1.155 & -0.123 & 0.135 & -0.248 & -0.912 & -0.496 & -0.909
            & 0.024  & 0.113 & -0.105 & 0.209  & 0.225  & 0.264  & 0.181  & 0.155 & -0.160 & 1.169  & 1.130  & 1.156 \\
SAC         & -0.140 & 0.090 & -0.207 & -1.554 & -0.676 & -1.635 & -0.167 & 0.081 & -0.207 & -2.057 & -0.807 & -2.219 
            & 0.140  & 0.111 & -0.069 & 1.263  & 2.011  & 1.242  & 0.162  & 0.139 & -0.107 & 1.165  & 1.518  & 1.154 \\
FactorVAE   & -0.048 & 0.134 & -0.175 & -0.335 & -0.271 & -0.348 & 0.006  & 0.127 & -0.147 & 0.047  & 0.041  & 0.112 
            & 0.160  & 0.142 & -0.132 & 1.128  & 1.211  & 1.013  & 0.356  & 0.159 & -0.119 & 2.234  & 2.995  & 1.907 \\  \hline
AlphaStock  & -0.164 & 0.153 & -0.245 & -1.072 & -0.669 & -1.098 & -0.017 & 0.148 & -0.166 & 0.115  & -0.102 & -0.043 
            & 0.122  & 0.140 & -0.126 & 0.871  & 0.968  & 0.892  & 0.372  & 0.178 & -0.134 & 1.781  & 2.776  & 1.869 \\
DeepPocket  & -0.036 & 0.135 & -0.175 & -0.270 & -0.207 & -0.258 & 0.006  & 0.127 & -0.148 & 0.050  & 0.043  & 0.115 
            & 0.165  & 0.142 & -0.126 & 1.165  & 1.311  & 1.045  & 0.346  & 0.157 & -0.116 & 2.197  & 2.971  & 1.882\\
DeepTrader  & -0.122 & 0.147 & -0.229 & -0.828 & -0.533 & -0.876 & 0.055  & 0.168 & -0.141 & 0.324  & 0.388  & 0.370 
            & 0.295  & 0.180 & -0.181 & 1.635  & 1.628  & 1.425  & 0.716  & 0.248 & -0.138 & 2.890  & 5.169  & 2.306 \\
THGNN       & -0.015 & 0.172 & -0.152 & -0.088 & -0.100 & -0.003 & 0.048  & 0.128 & -0.141 & 0.375  & 0.340  & 0.432
            & 0.271  & 0.141 & -0.094 & 1.921  & 2.871  & 1.778  & 0.644  & 0.204 & -0.146 & 3.147  & 3.414  & 2.543 \\
\textbf{LSR-IGRU} 
            & \textbf{0.192} & 0.187  & -0.168 & \textbf{1.026} & \textbf{1.143} & \textbf{1.009} 
            & \textbf{0.330} & 0.203  & -0.198 & \textbf{1.626} & \textbf{1.663} & \textbf{1.382}
            & \textbf{0.506} & 0.177  & -0.093 & \textbf{2.863} & \textbf{5.453} & \textbf{2.450} 
            & \textbf{0.835} & 0.222 & -0.109 & \textbf{3.747} & \textbf{7.604} & \textbf{2.914} \\  \hline
\end{tabular}}
\end{table*}

\subsection{Improved GRU Network}
Compared to the traditional GRU model, we make improvements in input design. While the conventional GRU uses daily time-series features of stocks as inputs at each step, this paper employs the obtained long-term relationship representations, short-term relationship representations, and time-series representations as inputs for each step of the GRU. Such input design takes into account the influences of both long-term and short-term relationships on the dynamics of stocks, rendering the model more adaptable and flexible. This approach allows the long-term and short-term relationships of stocks to affect the representations of relationships and time series of the next day during the training process of the GRU model. By comprehensively considering relationship information at different time scales, the model can more comprehensively capture the dynamic changes and correlations in the stock market, thereby improving prediction accuracy and robustness.

We already possess the original representations $x_{it}$, the long short-term representations  $x_{it}^{long''}$ and $x_{it}^{short''}$. The improved GRU model integrates these representations using the following formula:
\begin{equation}
\centering
\begin{split}
z_{it} = \sigma (W_{z}(x_{it}||x_{it}^{long''}||x_{it}^{short''}) + U_{z}h_{i(t-1)}+b_{z})  \\
r_{it} = \sigma (W_{r}(x_{it}||x_{it}^{long''}||x_{it}^{short''}) + U_{r}h_{i(t-1)}+b_{r})  \\
\widetilde{h_{it}}  = \tanh (W_{h}(x_{it}||x_{it}^{long''}||x_{it}^{short''}) + U_{h}(r_{it}\odot h_{i(t-1)})+b_{h}) \\
h_{it} = (1-z_{it})\odot \widetilde{h_{it}} + z_{it} \odot h_{i(t-1)}
\end{split}
\centering
\end{equation}
where $W_{z} \in \mathbb{R}^{H \times(F+2F')}$, $U_{z} \in \mathbb{R}^{H \times H}$ and $b_{z} \in \mathbb{R}^{H}$ are the weight parameters and bias parameters of the update gate, 
$W_{r} \in \mathbb{R}^{H \times(F+2F')}$, $U_{r} \in \mathbb{R}^{H \times H}$ and $b_{r} \in \mathbb{R}^{H}$ are the weight parameters and bias parameters of the reset gate, 
$W_{h} \in \mathbb{R}^{H \times(F+2F')}$, $U_{h} \in \mathbb{R}^{H \times H}$ and $b_{h} \in \mathbb{R}^{H}$ are the weight parameters and bias parameters used to calculate the candidate hidden state, 
$h_{i(t-1)} \in \mathbb{R}^{H}$ represents the hidden state of the previous time step, $\sigma$ is the activation function sigmoid \cite{han1995sig}, $\odot$ represents the element product operator, $||$ represents the connection operation.
$z_{it} \in \mathbb{R}^{H}$, $r_{it} \in \mathbb{R}^{H}$, $\widetilde{h_{it}} \in \mathbb{R}^{H}$, $h_{it} \in \mathbb{R}^{H}$ and $H$ is the dimension of the hidden state vector.

\subsection{Optimization Objectives}
After obtaining the time series representation and the long-short term relationship representation $h_{it}$, we input it into a multilayer perceptron for dimensionality reduction to obtain the predicted values. Finally, we compare the predicted values with the actual 
labels to calculate the loss, which is shown as follows:
\begin{equation}
\centering
\begin{split}
h_{it}^{'} = MLP(h_{it})  , \ \ \
& Loss(y,h_{it}^{'}) = \frac{1}{d'} {\textstyle \sum_{i=1}^{d'} (y - h_{it}^{'})^{2} } 
\end{split}
\centering
\end{equation}
where $d'$ is the number of samples per day, $y$ is the true value of the $i$-th sample on day $t$, and $h_{it}^{'}$ is the corresponding predicted value, $Loss()$ is the loss function mean square error.

\section{Experiments}
\subsection{Experimental Setttings}
\subsubsection{\textbf{Datasets}}
We conduct experiments using data from both the Chinese and American stock markets. These datasets include CSI 300, CSI 500, S\&P 500, and NASDAQ 100. For these four stock datasets, we use indicators such as opening price, closing price, highest price, lowest price, trading volume, and turnover. To ensure data consistency and mitigate the impact of outliers, we perform outlier removal and normalization on the data. We calculate the daily returns for each stock, defined as the percentage difference between the current day's closing price and the previous day, and rank the daily returns to serve as labels for the dataset.
We divide the data into training set (from 2018 to 2021), validation set (2022), and testing set (2023) in chronological order.

\subsubsection{\textbf{Baseline Models}}
We compare our LSR-IGRU model with the following baselines: 
BLSW \cite{BLSW}, CGM \cite{CGM}, LSTM \cite{LSTM}, ALSTM \cite{ALSTM}, GRU \cite{GRU}, Transformer \cite{Trans}, TRA \cite{TRA}, CTTS \cite{CTTS}, A2C \cite{A2C}, DDPG \cite{DDPG}, PPO \cite{PPO}, TD3 \cite{TD3}, SAC \cite{SAC}, AlphaStock \cite{AlphaStock}, DeepPocket \cite{DeepPocket}, DeepTrader \cite{DeepTrader}, FactorVAE \cite{FactorVAE} and THGNN \cite{THGNN}.

\subsubsection{\textbf{Parameter Settings}}
We set the time window $t$ to the past 15 days of historical data as training input. We utilize a two-layer graph attention network to encapsulate long and short-term relationship representations. The configuration of these layers is as follows: the first layer consists of 30 neurons, while the second layer consists of 15 neurons. 
In the GRU module, we implement two layers: the first layer consists of 15 neurons, followed by the second layer with 10 neurons. These layers culminate in a single output unit employed for forecasting future stock trends. 
During model training, we employ a batch size of 128 training sets, employing the mean squared error loss function, and optimize using the Adam \cite{kingma2014adam} optimizer with an initial learning rate of 0.0002. 
Each trading day, we construct a virtual investment portfolio, containing the top 10 stocks selected based on prediction scores.

\subsubsection{\textbf{Trading Protocols}}
Building upon \cite{dixon2017classification}, we employ a buy-hold-sell (BHS) trading strategy to assess the performance of stock trend prediction methods in terms of returns. During the testing period, on each trading day, we simulate stock traders to execute the following predictive trading process:
First, at the close of trading day $t$, traders utilize the model to compute prediction scores and rank the predicted returns for each stock.
Second, at the opening of the trading day $t+1$, traders sell the stocks bought on the day $t$ and purchase stocks with higher expected returns, particularly those ranked in the top $k$. It is worth noting that if a stock is consistently rated as having the highest expected returns, traders would continue to hold onto it.
Third, the impact of transaction costs is not considered in the experiment.

\begin{figure*}[t]
\centering
\subfigure[Time step of the CSI 500.]{\label{figure2_1}
\includegraphics[width=0.24\textwidth,height=0.15\textwidth]{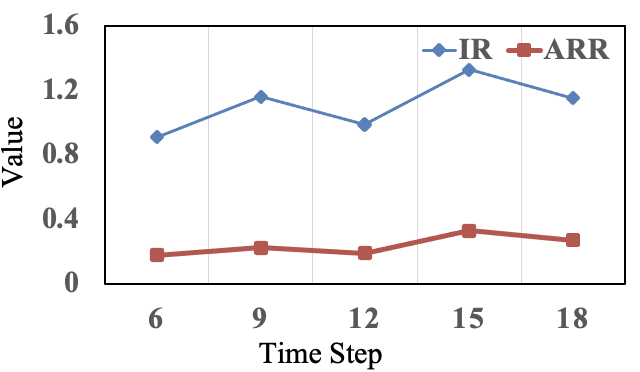}}
\subfigure[Train epoch of the CSI 500.]{\label{figure2_2}
\includegraphics[width=0.24\textwidth,height=0.15\textwidth]{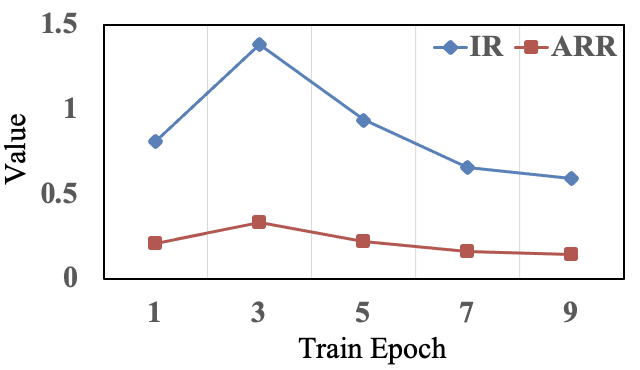}}
\subfigure[GCN layer's number of the CSI 500.]{\label{figure2_3}
\includegraphics[width=0.24\textwidth,height=0.15\textwidth]{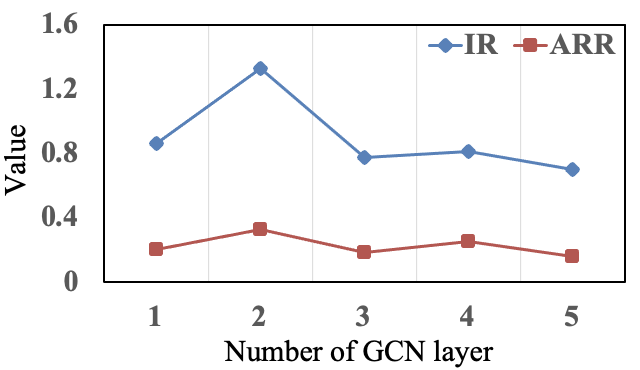}}
\subfigure[GRU layer's number of the CSI 500.]{\label{figure2_4}
\includegraphics[width=0.24\textwidth,height=0.15\textwidth]{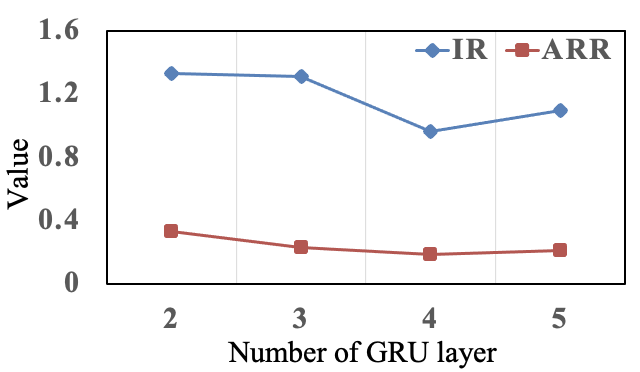}}
\subfigure[Time step of the NASDAQ 100.]{\label{figure2_5}
\includegraphics[width=0.24\textwidth,height=0.15\textwidth]{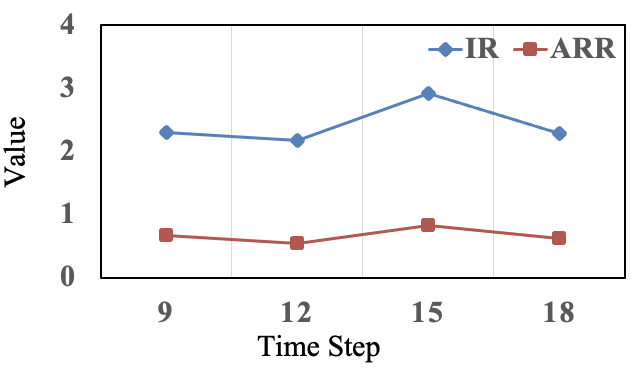}}
\subfigure[Train epoch of the NASDAQ 100.]{\label{figure2_6}
\includegraphics[width=0.24\textwidth,height=0.15\textwidth]{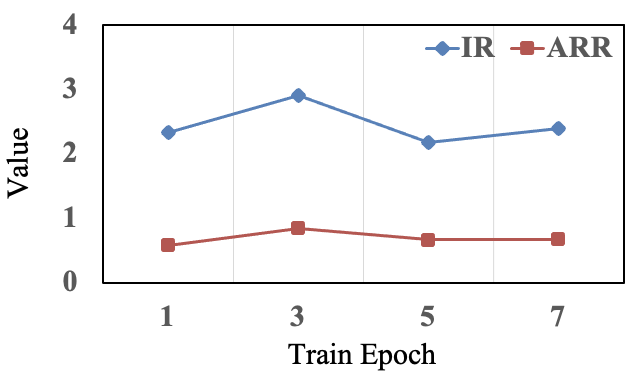}}
\subfigure[GCN layer's number of the NASDAQ 100.]{\label{figure2_7}
\includegraphics[width=0.24\textwidth,height=0.15\textwidth]{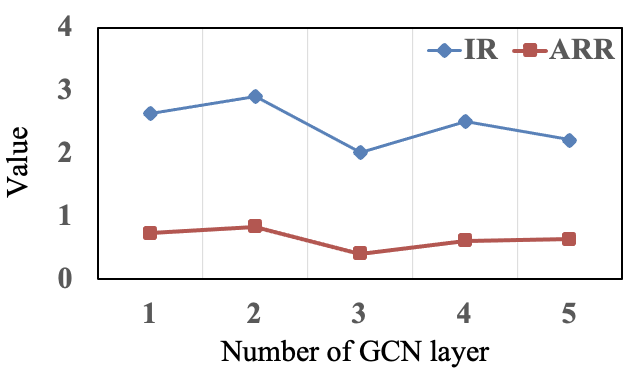}}
\subfigure[GRU layer's number of the NASDAQ 100.]{\label{figure2_8}
\includegraphics[width=0.24\textwidth,height=0.15\textwidth]{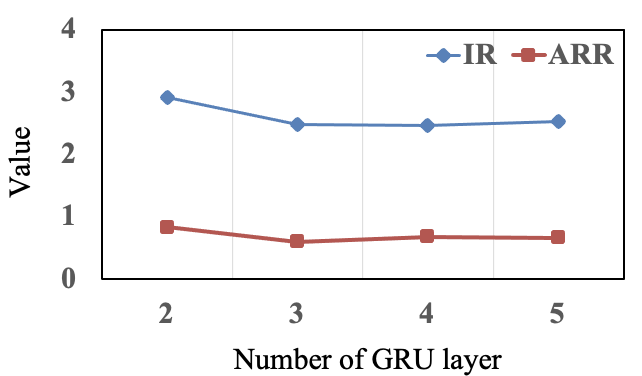}}
\caption{The results of parameter sensitivity experiments.}
\label{figure2}
\end{figure*}

\subsubsection{\textbf{Evaluation Metrics}}
Our goal is to accurately select stocks with the highest returns and we use six key financial indicators to evaluate the performance of both the baseline model and our proposed model.
Annualized Return Rate (ARR) is our primary metric, reflecting the effectiveness of the stock investment strategy by aggregating the daily returns of selected stocks over one year.
Annualized Volatility (AVoL) measures the standard deviation of daily returns, calculated by $AVoL=std(\frac{P_{t}}{P_{t-1}}-1 ) \times \sqrt{252}$, where $P_{t}$ and $P_{t-1}$ represent the stock price on day $t$ and day $t-1$ respectively.
Maximum Drawdown (MDD) captures the most significant decline from peak to trough, reflecting the maximum loss observed during the testing period.
Annualized Sharpe Ratio (ASR) evaluates the return per unit of volatility, given by  $ASR=\frac{ARR}{AVoL}$, while Calmar Ratio (CR) assesses the return per unit of drawdown, calculated as $CR=\frac{ARR}{|MDD|}$
Finally, the Information Ratio (IR) quantifies the excess return per unit of additional risk.
Together, these metrics provide a comprehensive framework for evaluating the performance and risk of our investment strategy. We prioritize lower AVoL and MDD values along with higher ARR, ASR, CR, and IR values, indicating superior performance.
To ensure robustness, we repeat each test ten times and report the average performance to mitigate variations from different initializations.

\subsection{Experimental Results}
In this section,  we conduct a comprehensive evaluation of various models' performance across different datasets, as shown in Tables \ref{table1}. The first seven rows of the tables cover traditional and deep learning models, including BLSW, CGM, GRU, LSTM, ALSTM, and Transformer, all of which are non-graph-based models. Despite demonstrating average performance across the datasets, these models' results imply that models not utilizing relational data struggle to reach optimal levels.
Subsequently, we introduce a series of more complex machine learning models and reinforcement learning models, such as TRA, CTTS, A2C, DDPG, PPO, TDC, SAC, and FactorVAE. While these models show improvement compared to the former, they still fell short of achieving optimal performance. Particularly, PPO, for instance, achieved the lowest AVol and MDD across the four datasets but exhibited mediocre performance in terms of ARR and IR, which are crucial metrics.
Further examination is conducted on graph-based models such as AlphaStock, DeepPocket, DeepTrader, and THGNN, which demonstrated significantly superior performance. This underscores the importance of relational information in enhancing model performance. Analysis based on these experimental results indicates that leveraging relational information can improve and strengthen the performance of stock prediction models. The model proposed in this paper fully exploits both the long-term invariant relationships and short-term dynamic relationships of stocks, combined with temporal information for prediction, thus achieving optimal performance. Observing Tables \ref{table1}, it is evident that LSR-IGRU achieves the highest ARR, ASR, CR, and IR across all four datasets, demonstrating the superiority of long short-term relationships in financial time series prediction.

\subsection{Parameter Sensitivity}
Figure \ref{figure2} shows the results of parameter sensitivity experiments on the CSI 500 and NASDAQ 100 datasets.
Figures \ref{figure2_1} and \ref{figure2_5} show ARR and IR increasing with the time step up to 15 days, then declining. This indicates the model effectively uses historical data for feature extraction. However, too little data provides insufficient information, while too much can introduce irrelevant details, reducing accuracy.
In Figures \ref{figure2_2} and \ref{figure2_6}, maximum ARR and IR are achieved at 3 training iterations. Performance decreases with increasing iterations, possibly due to overfitting or memorization of noise in the training set.
In Figures \ref{figure2_3} and \ref{figure2_7}, performance gradually improves with increasing GCN layers, peaking at the 2nd layer. However, further layers lead to performance decline, suggesting a balance between complexity and generalization.
Figures \ref{figure2_4} and \ref{figure2_8} show initial performance improvement with increasing GRU layers, but overfitting occurs beyond a certain threshold, resulting in poor test set performance.

\begin{table}[ht]
\centering
\caption{Performance evaluation of ablated models for financial time series prediction in CSI 300 and S\&P 500 datasets.}
\label{table2}
\footnotesize
\setlength{\tabcolsep}{0mm}{
\begin{tabular}{c|cccccc|cccccc}
\hline
Datasets& \multicolumn{6}{c|}{CSI 300}& \multicolumn{6}{c}{S\&P 500}\\ 
Model & ARR $\uparrow$ & AVol $\downarrow$ & MDD $\downarrow$ & ASR $\uparrow$ & CR $\uparrow$ & IR $\uparrow$ & ARR $\uparrow$ & AVol $\downarrow$ & MDD $\downarrow$ & ASR $\uparrow$ & CR $\uparrow$ & IR $\uparrow$ \\ \hline
I & -0.229 & \textbf{0.156} & -0.290 & -1.469 & -0.790 & -1.631 & 0.324 & 0.169 & -0.139 & 1.917 & 2.331 & 1.657 \\
II  & 0.079 & 0.167 & \textbf{-0.112} & 0.473 & 0.704 & 0.521 & 0.356 & 0.245 & -0.210 & 1.450 & 1.692 & 1.372  \\
III & 0.111 & 0.189 & -0.172 & 0.585 & 0.645 & 0.619 & 0.307 & 0.169 & -0.136 & 1.825 & 2.260 & 1.697  \\ \hline
I+II & 0.157 & 0.163 & -0.164 & 0.966 & 0.958 & 0.948 & 0.401 & 0.177 & -0.137 & 2.267 & 2.929 & 1.993 \\
I+III & 0.173 & 0.181 & -0.169 & 0.954 & 1.022 & 0.930 & 0.412 & \textbf{0.165} & -0.120 & 2.489 & 3.419 & 2.198 \\
II+III & 0.145 & 0.286 & -0.181 & 0.506 & 0.796 & 0.632 & 0.439 & 0.195 & -0.144 & 2.252 & 3.047 & 1.982 \\  \hline
\textbf{All} & \textbf{0.192} & 0.187 & -0.168 & \textbf{1.026} & \textbf{1.143} & \textbf{1.009}& \textbf{0.506} & 0.177 &\textbf{ -0.093} & \textbf{2.863} & \textbf{5.453} & \textbf{2.450} \\
\hline
\end{tabular}}
\end{table}

\subsection{Ablation Study}
We conduct comprehensive disintegration experiments to assess the model and investigate the impact of removing individual submodules on its performance. Employing the CSI 300 and S\&P 500 datasets due to space constraints, we present detailed results in Table \ref{table2}. Our model comprises three primary submodules: the GRU module, the long-term relationship representation module, and the short-term relationship representation module. As shown in Table \ref{table2}, using any single module alone failed to achieve the optimal performance level, while combining any two modules exhibited stronger predictive accuracy. This indicates that each module can contribute to enhancing the model's performance to a certain extent. Of particular note is that when all submodules are integrated simultaneously, our model demonstrates the best performance, further confirming the importance of the representation of long and short relationships in our model. This representation not only enables the full exploration of complex correlations between stocks but also significantly improves predictive accuracy. Our study emphasizes the collaborative and complementary roles among different modules during the model integration process, further underscoring the importance of integrating multiple information sources in enhancing predictive performance.

\begin{figure}[t]
\centering
\subfigure[CSI 300 strategy backtest performance.]{\label{figure4_1}
\includegraphics[width=0.48\textwidth,height=0.155\textwidth]{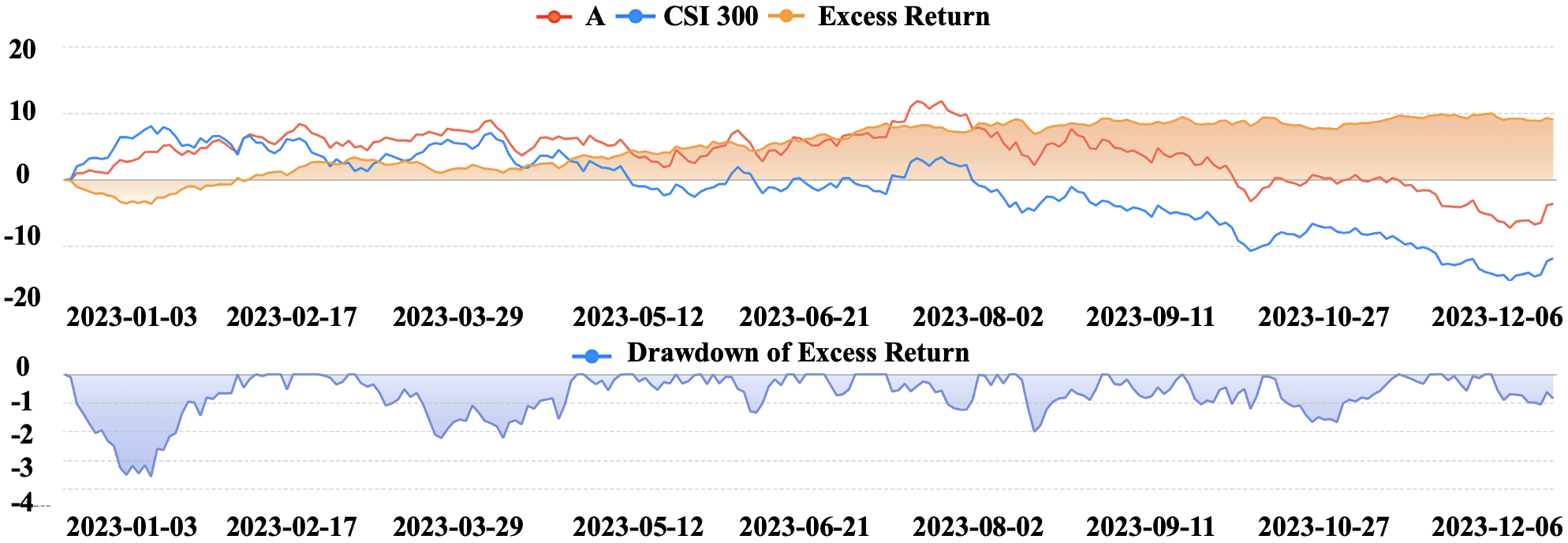}}
\subfigure[CSI 500 strategy backtest performance.]{\label{figure4_2}
\includegraphics[width=0.48\textwidth,height=0.155\textwidth]{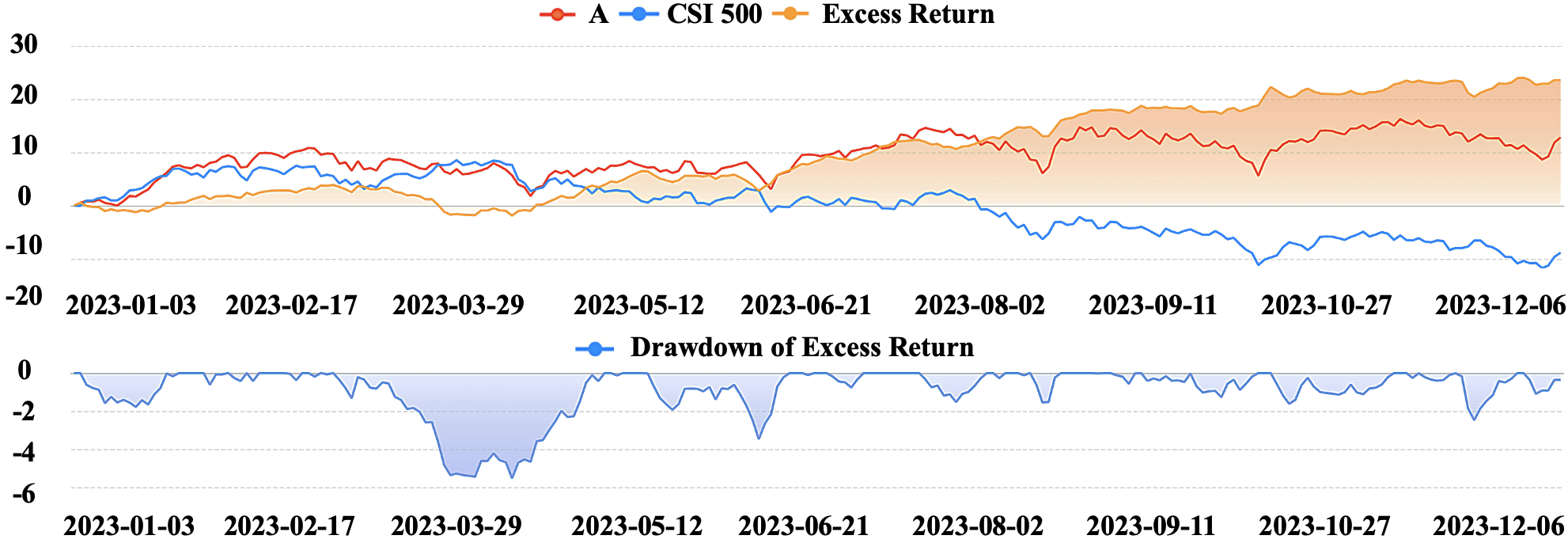}}
\caption{The performance of the strategy backtest.}
\label{figure4}
\end{figure}

\subsection{Deployment}
This section elaborates on how to apply our model to real-world trading systems and conduct strategy updates and transactions in the real algorithmic trading platform of EMoney Inc. Our model is trained every month, predicting trading data for each day after the end of trading. Based on the predictions, we employ different strategies for trading within the first half hour of the next trading day's opening. The strategies we utilize are based on the combined optimization of CSI 300 and CSI 500, each utilizing stock pools from CSI 300 and CSI 500 respectively. Figure \ref{figure4} illustrates the effects of different strategies. In Figure \ref{figure4_1} and \ref{figure4_2}, the red curve represents the absolute returns of the model, i.e., the actual returns of our model. The blue curve represents the returns of the Shanghai CSI 300 and CSI 500 indices, reflecting the overall market returns. The yellow curve represents the excess returns, i.e., the additional returns obtained relative to the market index by our model. Over a year, all strategies of our model significantly outperform the market. Additionally, the second part of Figure \ref{figure4} shows the excess return drawdown rate, which reflects the model's good risk management capabilities by measuring the extent to which excess returns decline from their peak to their lowest point. It can be observed from the figure that our model maintains a very low drawdown rate over a long period, reaching only about 5\% in the worst case. This indicates that our model not only pursues returns but also possesses the ability to prudently manage risks, making it outstanding in real trading markets.

\section{Conclusion}
In this paper, we introduce a novel stock prediction model called LSR-IGRU, which integrates long-term and short-term stock relationships with improved GRU inputs. Our model leverages stock industry information at the secondary level to construct a long-term relationship network for the first time, while also incorporating overnight price information to build a short-term relationship network. By enhancing the modeling capabilities of temporal information and long-term and short-term relationships in each step of the GRU model's input, we improve the accuracy of predicting stock trend changes. We conduct extensive experiments on four different datasets including CSI 300, CSI 500, S\&P 500, and NASDAQ 100. The experimental results demonstrate that the proposed method in this paper outperforms the state-of-the-art baseline models, proving the model's superior performance and generalization ability. Additionally, we deploy our model into the real algorithmic trading platform of the financial service provider EMoney Inc. In various combination optimization strategies, the model's cumulative returns significantly outperformed existing models, validating its effectiveness and feasibility in practical applications.

\begin{acks}
This work was supported by the National Key Research and Development Program of China under Grant 2022YFB4501704, the Shanghai Science and Technology Innovation Action Plan Project under Grant 22YS1400600 and 22511100700, the National Science Foundation of China under Grant 62222312, and the Fundamental Research Funds for the Central Universities.
\end{acks}

\bibliographystyle{ACM-Reference-Format}
\bibliography{cikm}

\end{document}